\documentclass[aps, prl, groupaddress, reprint, showpacs]{revtex4-1}

\usepackage{amsfonts}
\usepackage{amsmath}
\usepackage{amssymb, bm}
\usepackage{graphicx, epstopdf, color}
\usepackage{soul}
\usepackage{cleveref, appendix}

\begin{document}

\title{Current-temperature scaling for a Schottky interface with non-parabolic energy dispersion}

\author{Y. S. Ang}
\email{Corresponding author. yeesin\_ang@sutd.edu.sg}
\author{L. K. Ang}
\email{Corresponding author. ricky\_ang@sutd.edu.sg}
\affiliation{SUTD-MIT International Design Center, Singapore University of Technology and Design, Singapore 487372}
\affiliation{Engineering Product Development, Singapore University of Technology and Design, Singapore 487372}

\begin{abstract}
	 
	In this paper, we study the Schottky transport in narrow-gap semiconductor and few-layer graphene in which the energy dispersions are highly non-parabolic. 
	We propose that the contrasting current-temperature scaling relation of $J\propto T^2$ in the conventional Schottky interface and $J\propto T^3$ in graphene-based Schottky interface can be reconciled under Kane's $\mathbf{k} \cdot \mathbf{p}$ non-parabolic band model for narrow-gap semiconductor. 
	Our new model suggests a more general form of $J\propto \left(T^2 + \gamma k_BT^3 \right)$, where the non-parabolicty parameter, $\gamma$, provides a smooth transition from $T^2$ to $T^3$ scaling. 
	For few-layer graphene,	it is found that $N$-layers graphene with $ABC$-stacking follows $J\propto T^{2/N+1}$ while $ABA$-stacking follows a universal form of $J\propto T^3$ regardless of the number of layers. 
	Intriguingly, the Richardson constant extracted from the Arrhenius plot using an incorrect scaling relation disagrees with the actual value by two orders of magnitude, suggesting that correct models must be used in order to extract important properties for many novel Schottky devices.

\end{abstract}

\maketitle

\section{Introduction}

Translating the unusual physical properties of novel nanomaterial-based heterostructures into functional device applications has become one of the major research goals in recent years \cite{geim}. 
One important heterostructure is the metal/semiconductor interface, commonly known as the \emph{Schottky interface} \cite{xu_y}, where novel applications such as broadband ultrasensitive photodetector \cite{tsai}, gate-tunable Schottky barrier \cite{yang}, promising solar cell performance \cite{tsai1} and ultrafast phototransistor \cite{lee} have recently been demonstrated.
The current transport across a Schottky interface is mainly due to majority carriers. 
In general, there are three different transport mechanisms, namely diffusion of carriers from the semiconductor into the metal, thermionic emission of carriers across the Schottky barrier and quantum-mechanical tunneling through the barrier \cite{liang_WS}. 
For the thermionic emission, the Schottky diode equation is written as \cite{crowell}
\begin{equation}\label{sch}
J = \bar{J} \left( e^{\frac{eV}{\eta k_BT}} - 1 \right),
\end{equation}
where $\bar{J}$ is the reverse saturation current density determined by the thermionic emission process, $V$ is the bias voltage and $\eta$ is an ideality factor. 
For bulk materials with parabalic energy disperson ($E_k \propto k^2$), the reversed saturation current density $\bar{J}$ takes the well-known Richardson form of \cite{richardson, crowell2}
\begin{equation}\label{RD}
\bar{J}_{R} \propto T^2 e^{-\frac{\Phi}{k_BT}},
\end{equation} 
where $\Phi$ denotes the magnitude of the Schottky barrier's height.
The exponential term, $e^{-\Phi/k_BT}$, in Eq. (\ref{RD}) originates from the classical Boltzmann statistics and is universal regardless of the form of the transport electron energy dispersion while the $\bar{J}\propto T^2$ current-temperature scaling relation is a signature of the parabolic energy dispersion of the transport electrons. 

For novel materials with non-parabolic energy dispersion [see Figs. 1(a)-(d) for examples of non-parabolic energy dispersions], the validity of $\bar{J}\propto T^2$ should be verified. 
Although it is well-known that the energy dispersion plays an important role in governing the Schottky transport, the traditional $\bar{J}\propto T^2$ model is still widely used in the vast majority of recent experimental works on Schottky interfaces composed of novel materials, such as MoS$_2$, black phosphorus, graphene and few-layer graphene, where the dispersion is highly non-parabolic \cite{yang, chen, yim, kim, mohammed, shivaraman, tongay, an, yu2, das, tian, kang}. 
This highlights the need to reformulate the Schottky model in order to uncover the underlying physics in these structures.
For a monolayer graphene, it is recently reported that $\bar{J}$ has an unconventional form of $\bar{J}_{Dirac} \propto T^3 e^{-\Phi/k_BT}$ \cite{liang1}.
The $\bar{J}_{Dirac}\propto T^3$ behavior can be regarded as the \emph{Dirac-Schottky scaling relation} and is a signature of the linear energy dispersion in graphene \cite{neto}. 
As the form of the energy dispersion can crucially affect the scaling, the Schottky transport model has to be reformulated for Schottky interface made up of non-parabolic dispersions-based materials. 
The very distinct forms between the Schottky $T^2$-scaling and the Dirac-Schottky $T^3$-scaling also prompts us to investigate whether the $\bar{J}_{R}\propto T^2$ and the $\bar{J}_{Dirac}\propto T^3$ can be connected via a unique energy dispersion that is `intermediate' between parabolic and linear and whether graphene multilayer follows other forms of unconventional scaling relation.

In this paper, we study the Schottky transport in narrow-gap semiconductor and in few-layer graphene (FLG) in which the energy dispersion is highly non-parabolic. We show that the Schottky $T^2$-scaling and the Dirac-Schottky $T^3$-scaling can be unified under Kane's $\mathbf{k}\cdot \mathbf{p}$ band model for narrow-gap semiconductor in which the band non-parabolicity is captured by the \emph{non-parabolicity paramter}, $\gamma$ \cite{kane, askerov,conwell}. We obtained a \emph{Kane-Schottky scaling relation} of
\begin{equation}
\bar{J}_{Kane} \propto  \left( T^2 + 2\gamma k_B T^3 \right) e^{-\frac{\Phi}{k_BT}}. \label{kane}
\end{equation}
The scaling relation exhibits a mixture of $T^2$ and $T^3$. Here, $\gamma$ is responsible for the continuous transition from $\bar{J} \propto T^2$ to $\bar{J} \propto T^3$ scaling. 
In the case of perfectly parabolic dispersion ($\gamma \to 0$) and perfectly linear dispersion ($\gamma \to \infty$), the scaling becomes $\bar{J} \propto T^2$ and $\bar{J} \propto T^3$, respectively. 
Thus, Kane-Schottky scaling relation is a more general SR that connects the Schottky $T^2$-scaling and the Dirac-Schottky $T^3$-scaling. 
In few-layer graphene (FLG), we found that the scaling is strongly dependent on the stacking order. 
For $ABA$-stacked $N$-layer FLG, the Schottky current shows an $N$-fold enhancement due to the presence of $N$ conduction subbands. Peculiarly, $\bar{J}^{(N)}_{ABA}$ follows the Dirac-Schottky $T^3$-scaling universally regardless of the number of layers. This is in contrast to $ABC$-FLG where $\bar{J}^{(N)}_{ABC}$ follows $N$-dependent scaling of $\bar{J}^{(N)}_{ABC}\propto T^{2/N+1}$. 
Finally, we show that the \emph{Richardson constant} extracted from the Arrhenius plot disagrees with the actual values by 2 orders of magnitude when an incorrect $T^2$-scaling is used. 
This emphasizes the importance of using a correct model when interpreting the experimental data in Schottky devices of non-parabolic energy dispersions.

\section{Theory}

The Schottky transport model is shown in Fig. 1(d). $\bar{J}$ is determined by the thermionic emission process that we will briefly describe the formalism here \cite{fowler, liang_WS}. The energy of the emitted electron can be written as $E = E_\perp + E_\parallel$, where $E_\perp$ is the energy component along the emission $z$-direction and $E_\parallel$ is the energy component lies in the $xy$-plane [see inset of Fig. 1(d)]. The electron emission current density is given as 
\begin{equation}\label{eq1}
\bar{J} = \int_{\Phi}^{\infty} N(E_\perp)D(E_\perp)dE_\perp,
\end{equation}
where $D(E_\perp)$ is the transmission probability and $\Phi$ is the Schottky barrier. For over-barrier process, $D(E_\perp)$ can be approximated by $D(E_\perp) = \Theta(E_\perp - \Phi)$. The electron supply function $N(E_\perp)$ can be expressed as $N(E_\perp)dE_\perp = dE_\perp \int_{E_\perp}^\infty n(E,E_\perp) dE$ where the electron supply density is
\begin{equation}\label{eq2}
n(E,E_\perp) dE dE_\perp = g_{s,v} ev_\perp f(E)  \frac{d^3 k}{(2\pi)^3}.
\end{equation}
The group velocity component along the emission direction is given as $v_\perp = \hbar^{-1} dE_\perp / dk_\perp$ and $f(E)$ is the Fermi-Dirac distribution function. The $k$-space integration can be rewritten as $d^3k =  k_\parallel dk_\parallel d\phi dk_\perp$ where $\mathbf{k}_\parallel = (k_x,k_y)$ denotes the in-plane crystal momentum of the transport electron, $k_\perp$ denotes the out-of-plane momentum component and $\phi = \tan^{-1}k_y/k_x$. Since the energy of the over-barrier electron is much larger than the Fermi level, the Fermi Dirac distribution function can be approximated by the Boltzmann distribution function. We can simplify $n(E,E_\perp)dEdE_\perp$ as 
\begin{equation}\label{eq3}
n(E,E_\perp) dE dE_\perp = dE_\perp \frac{g_{s,v} e}{(2\pi)^3\hbar} f_{MB}(E)  k_\parallel dk_\parallel d\phi,
\end{equation} 
where $f_{MB}(E) $ is the Boltzmann distribution function. In order to complete the $\int (\cdots) dE$ integral in $N(E_\perp)dE_\perp$, the $k$-space differentials $k_\parallel dk_\parallel$ needs to be converted into an $E_\parallel$-space differentials. The $k_\parallel dk_\parallel \to dE_\parallel$ transformation depends on the actual form of the $E_\parallel$-$k_\parallel$ relation, i.e. the energy dispersion. Therefore, $N(E_\perp)dE_\perp$ contains all information of the energy dispersion and plays an important role in determining the form of the Schottky transport current.

\begin{figure}[t]
	\includegraphics[scale=.4]{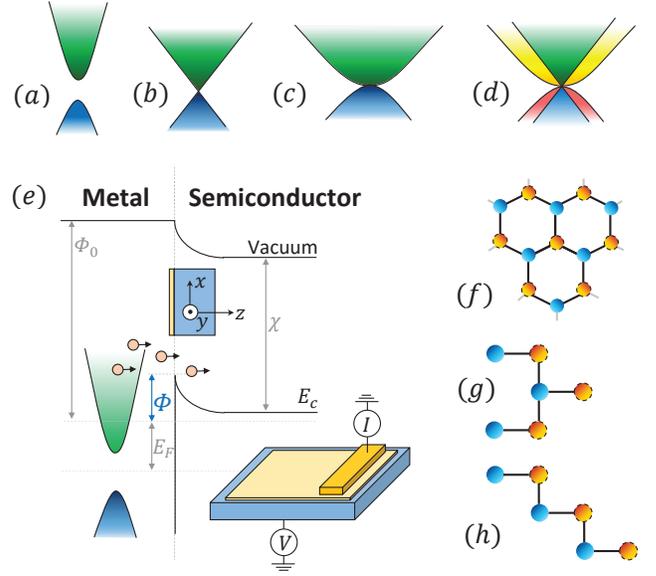}
	\caption{Non-parabolic energy dispersion and model of Schottky transport. (a) Kane's non-parabolic band; (b) low energy Dirac cone in graphene; low energy dispersion of trilayer graphene with (c) $ABA$-stacking and (d) $ABC$-stacking orders; (e) Schottky transport model in a metal/semiconductor interface. $\Phi_0$ is the work function of the metal, $\chi$ is the electron affinity of the semiconductor and the Schottky barrier is $\Phi = \Phi_0 - \chi$. Inset shows the coordinate system and the geometry of the Schottky interface studied in this work; (f) crystal structure of graphene layer. Solid (Dashed) circle denotes $A$-($B$-)sublattice; (g) $ABA$-stacking order; and (h) $ABC$-stacking order.}
\end{figure}

\section{Results and Discussions}

In this Section, we present the Schottky transport models for two classes of non-parabolic energy dispersions: (i) Kane's non-parabolic energy dispersion for narrow-gap semiconductor; and (ii) FLG with $ABA$-stacking and with $ABC$-stacking. For (i), we further consider two related band structure effects, i.e. Kane's model with band anisotropy and parabolic model with higher order $k^4$ correction term. The details of the derivation is presented in the Appendices. 

\subsection{Kane-Schottky transport model}

The electron transport in narrow gap semiconductor is well-described by Kane's non-parabolic band model \cite{kane, askerov, wu}. As the parabolic energy band is only a good approximation near the conduction band edge, Kane's model is also an improved band model especially for higher energy transport electrons \cite{jacoboni, ruf, anile}. Kane's non-parabolic energy dispersion is given as $E_\parallel(1 + \gamma E_\parallel) = \hbar^2 k_\parallel^2/2m$ where $\gamma = (1 - m/m_0)/E_g$ denotes the non-parabolicity of the dispersion, $m_0$ is the bare electron mass and $E_g$ is the magnitude of the bandgap. $\gamma k_B$ typically lies in the range of $10^{-4}$ K$^{-1}$ to $10^{-3}$K$^{-1}$ for sub-eV narrow-gap semiconductor such as PbSe, InAs, InSb and the topological insulators HgCdTe and Bi$_2$Te$_3$ \cite{pryor, martinez, hansen, yavorsky}. The energy dispersion can be re-expressed as $E_\parallel = (2\gamma)^{-1}\left(\sqrt{1+4\gamma\hbar^2 k_\parallel^2/2m } -1\right)$. For small $\gamma$, we recover the parabolic dispersion $E_\parallel \propto k_\parallel^2 $.  
For large $\gamma$, $E_\parallel (1+\gamma E_\parallel) \approx \gamma E_\parallel^2$ and this yields a linear dispersion of $E_\parallel \propto k_\parallel $. 
Hence, Kane's model connects the two extreme cases of perfectly parabolic and perfectly linear dispersion via $\gamma$. 
Solving Eqs. (\ref{eq1}), (\ref{eq2}) and (\ref{eq3}) using Kane's non-parabolic energy dispersion, we obtained the \emph{Kane-Schottky diode equation} as 
\begin{equation}\label{uni}
	J_{Kane} = \frac{g_{s,v} emk_B^2}{4\pi^2\hbar^3 } \left(T^2 + 2\gamma k_BT^3\right) e^{-\frac{\Phi}{k_BT}} \left(e^{\frac{eV}{\eta k_BT} - 1}\right).
\end{equation}
The detailed derivation can be found in the Appendix A.
The reverse saturation current density exhibits a combination of the Schottky $T^2$-scaling and the Dirac-Schottky $T^3$-scaling, i.e. $J\propto \left( T^2 + 2\gamma k_BT^3 \right)$.
This finding concludes that the Kane-Schottky model gives a more general scaling relation as it unifies both Schottky and Dirac-Schottky scaling relations via $\gamma$. 
For highly parabolic limit, $\gamma \to 0$ and this yields the conventional Schottky $T^2$-scaling with reverse saturation current density $\bar{J}_{\gamma \to 0} = \mathcal{A} T^2 e^{-\frac{\Phi}{k_BT}}$ where $\mathcal{A} = g_{s,v}emk_B^2/4\pi^2\hbar^3 $ is the Richardson constant. 
In the extremely non-parabolic limit of $\gamma \to \infty$ (i.e. perfectly linear dispersion), the energy dispersion becomes linear in $k$, i.e. $E_\parallel = \hbar v_F k$ where $v_F \equiv \sqrt{1/2m\gamma}$ and the $T^3$ term dominates. In this case, Eq. (\ref{uni}) reduces to the Dirac-Schottky form of $J = \mathcal{B} T^3 e^{-\frac{\Phi}{k_BT}}$ \cite{liang1} where $\mathcal{B} =g_{s,v}ek_B^3/4\pi^2\hbar^3 v_F^2$ is the modified Richardson constant in graphene. 

The Kane-Schottky diode model has an implication in the experimental determination of the Richardson constant. In Fig. 2, we generate the Kane-Schottky current density, $J$, with $\Phi= 0.30$ eV and $\eta = 1.1$ for temperature from $200$ K to $500$ K and plotted the $1/T$-Arrhenius plot using different scaling relations of: (i) $\log[J/(T^2 + 2\gamma k_BT^3)]$; (ii) $\log(J/T^2)$; and (iii) $\log(J/2\gamma k_BT^3)$. Due to the dominating $\exp(\Phi/k_BT)$, scaling (ii) and (iii) are both well-fitted by straight lines. The Schottky barrier's heights extracted from the gradients of the linear-fit are $\Phi = (0.31, 0.28)$ eV respectively for (ii) and (iii). This deviates only slightly from the actual value of $0.30$ eV. However, the Richardson constant, determined from the $y$-intercepts of the linear-fit, is $\mathcal{A}_{\text{fit}} = (2.46, 0.0028)\mathcal{A}_0$ respectively for scaling (ii) and (iii) and disagrees significantly with the actual value $\mathcal{A}_0 \equiv eg_{s,v}mk_B^2/4\pi^2\hbar^3$. This illustrates the importance of using the correct scaling in the Arrhenius plot, instead of assuming the conventional $T^2$-scaling, when extracting $\mathcal{A}_{\text{fit}}$ from experimental data.

\begin{figure}[t]
	\includegraphics[scale=.9]{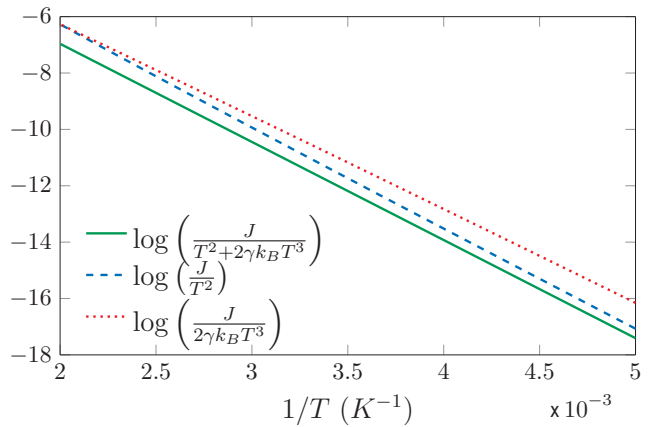}
	\caption{Arrhenius plot of the Kane-Schottky model using different scaling relation with $\gamma k_B = 10^{-3}$ K$^{-1}$. Kane-Schotkky, Schottky, and Dirac-Schottky scaling relations are denoted by solid, dashed and dotted lines, respectively. The reverse-bias is set to $V = -1$ V to ensure current saturation.}
\end{figure}

For completeness, we further demonstrate that the Kane-Schottky scaling relation is robust against band anisotropy and can be similarly obtained by including a higher order $k^4$ correction term in the parabolic dispersion. For the former case, we have $E_\parallel(1+\gamma E_\parallel) = \hbar^2 k_x^2/2m_x + \hbar^2 k_y^2/2m_y$ where $(m_x,m_y)$ is the anisotropic effective mass in $x$-and $y$-direction respectively. This yields a Kane-Schottky scaling in the form of
\begin{equation}\label{uni_an}	
\bar{J}_{m_x\neq m_y} =  \frac{g_{s,v}ek_B^2\sqrt{m_xm_y}}{4\pi^2 \hbar^3} \left( T^2 + 2k_B\gamma T^3 \right) e^{-\frac{\Phi}{k_BT}} .
\end{equation}
The latter case represents an alternative approach to account for the band non-parabolicity via a higher-order $k^4$ term, i.e. $E= \alpha k^2 - \beta k^4$ where $\alpha \equiv \hbar^2/2m$ and $\beta$ is a small correction factor. In this case, we obtain 
\begin{equation}
\bar{J}_{\alpha k^2 -\beta k^4} = \frac{g_{s,v}e}{8\pi^2\hbar\sqrt{\beta}} \left(k_BT\right)^{3/2} \mathcal{D}_+\left(\sqrt{\frac{\varepsilon_0}{k_BT}}\right)e^{-\frac{\Phi}{k_BT}},
\end{equation} 
where $\mathcal{D}_+(x) = e^{-x^2} \int^x_0 e^{t^2}$ is the Dawson integral and  $\varepsilon_0 = \alpha^2/4\beta$ is an characteristic energy. Note that as $\beta \ll \alpha$, $\varepsilon \gg k_BT$ for all practical temperature. Using the fact that $\mathcal{D}_+(x) \approx 1/2x + 1/4x^3$ for large $x$, we obtain the Kane-Schottky scaling of
\begin{equation}\label{k4}
\bar{J}_{\alpha k^2 -\beta k^4} \approx \frac{g_{s,v}em^*k_B^2}{4\pi^2\hbar^3}\left[T^2 + \frac{8m^2\beta}{\hbar^4}k_BT^3\right] e^{-\frac{\Phi}{k_BT}}.
\end{equation}

\subsection{Few-layer graphene Schottky transport model}

The electronic properties of FLG is sensitively dependent on the number of layer, $N$, and the stacking order \cite{charlier, yuan, lui, mak, bao, zhu, guinea1, guinea2, henni, kim1, jhang, craciun, grushina}. FLG with $ABA$-and $ABC$-stacking are the most thermodynamically stable stacking orders \cite{aoki}. The energy dispersion of both stacking orders are highly non-parabolic and this motivates us to develop a non-parabolic Schottky transport model for $ABA$-and $ABC$-FLG-based Schottky interfaces (see Appendix B for detailed derivations). Ignoring the layer-asymmetry bandgap \cite{mccann, koshino2}, the energy dispersion of the $n$-subband of $ABA$-FLG is  \cite{guinea2, min, koshino}
\begin{equation}
	E_{k_\parallel,n} = t_\perp \cos{\left(\frac{\pi n}{N+1}\right)} \pm \sqrt{\left(v_Fk_\parallel\right)^2 + t_\perp^2 \cos^2{\left(\frac{\pi n}{N+1}\right)}},
\end{equation}
where $N \geq 3$ is the number of layers, $t_\perp \approx 0.39$ eV is the interlayer hopping parameter \cite{gruneis, aoki, guinea2}, $v_F = 10^6$ m/s is the Fermi velocity and $n=1,2, \cdots, N$ represents each of the $2N$ subbands. $\bar{J}$ can be derived as 
\begin{equation}\label{AB}
	\bar{J}_{ABA}^{(N)} = N \times \frac{eg_{s,v} k_B^3}{ 4\pi^2 \hbar^3 v_F^2 } T^3 e^{-\frac{\Phi}{k_BT}}.
\end{equation}
The Schottky current exhibits an $N$-fold enhancement and a universal Dirac-Schottky $T^3$ scaling for all $N$. The $N$-fold enhancement can be explained by the presence of $N$ conduction subbands \cite{gruneis}. The $N$-independent $T^3$ scaling is a rather surprising result. As the $ABA$-FLG contains multiple non-parabolic subbands, one would expect a mixture of $T^2$ and $T^3$ terms in the Schottky current equation. However, we found that the $T^2$ term generated by $j<N$ subband is exactly canceled out by that of the $(N-j)$ subband (where $j \neq N/2$ is a positive integer). This mutual cancellation leads to the universal Dirac-Schottky $T^3$ scaling in $ABA$-FLG regardless of the number of layers, $N$.

\begin{figure}[t]
	\includegraphics[scale=.7]{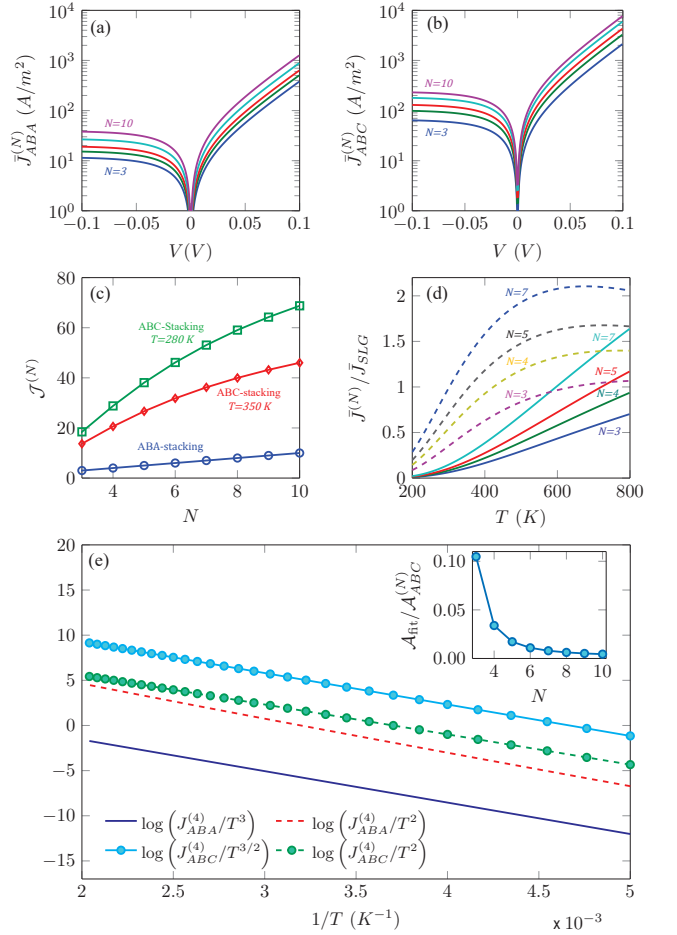}
	\caption{Schottky transport in few-layer graphene with $ABA$-and $ABC$-stacking orders. $J$-$V$ characteristics of (a) $ABA$-stacking; and (b) $ABC$-stacking at $T=300$ K and $\eta = 1.1$ with for $N=(3,4,5,7,10)$; (c)the $N$-dependence of the normalized Schottky current; (d) the temperature dependence of $\bar{J}^{(N)}/\bar{J}_{Dirac}$. Solid (Dashed) lines denote $ABA$-($ABC$-)stacking; (e) Arrhenius plot of tetralayer graphene ($N=4$). The inset shows the $N$-dependence of $\mathcal{A}_{\text{fit}}/\mathcal{A}^{(N)}_{ABC}$.}
\end{figure}

In the case of $ABC$-FLG, the low energy two-band effective tight-binding model \cite{guinea1, min, zhang_F} gives an energy dispersion of $E_\parallel = \left( \hbar v_F \right)^N / t_\perp^{ N-1 } k_\parallel^N$. We obtain
\begin{equation}\label{ABC}
	\bar{J}_{ABC}^{(N)} = \frac{\Gamma(2/N)}{N} \left( \frac{t_\perp}{k_B}  \right)^{2-\frac{2}{N}} \frac{eg_{s,v} k_B^3}{4\pi^2 \hbar^3 v_F^2 }  T^{ \frac{2}{N} + 1 }  e^{-\frac{\Phi}{k_BT}},
\end{equation}
where $\Gamma(x)$ is a gamma function. The Schottky current follows an $N$-dependent scaling relation of $\bar{J}_{ABC}^{(N)} \propto T^{2/N+1}$ in contrast to $ABA$-FLG. The $J$-$V$ characteristics of $ABA$-FLG and $ABC$-FLG are plotted, respectively, in Fig. 3(a) and Fig. 3(b) for a typical Schottky barrier height of $\Phi = 0.5$ eV. In general, the Schottky current increases with $N$ in both stacking orders and $J_{ABC}^{(N)}$ is about an order of magnitude larger than $J_{ABA}^{(N)}$. In Fig. 3(c), we plot the layer-dependence of the normalized Schottky current, i.e. $\mathcal{J}_i \equiv \bar{J}^{(N)}_{i}(T)/\tilde{J}_0(T)$ for $i=(ABA,ABC)$ where $\tilde{J}_0(T) \equiv eg_{s,v} k_B^3T^3 e^{-\Phi/k_BT}/(4\pi^2\hbar^3v_F^2)$. $\mathcal{J}_{AB}$ and $\mathcal{J}_{ABC}$ exhibits distinct forms of $N$-dependence. For $ABA$-FLG, $\mathcal{J}_{ABA} \propto N$ and is temperature-independent. In contrast, $\mathcal{J}_{ABA}$ exhibits a \emph{temperature-dependent} nonlinear growth with $N$. To compare the Schottky transport of FLG with that of the monolayer graphene, we define the following ratio:
\begin{subequations}\label{ratio}
\begin{align}
\frac{\bar{J}^{(N)}_{ABA}}{\bar{J}_{Dirac}} &= N e^{-\frac{\Delta \Phi^{(N)}_{AB} }{k_BT}} \label{ratio1};\\
\frac{\bar{J}^{(N)}_{ABC}}{\bar{J}_{Dirac}} &= \frac{\Gamma(2/N)}{N} \left(\frac{t_\perp}{k_BT}\right)^{2-\frac{2}{N}} e^{-\frac{\Delta \Phi^{(N)}_{ABC} }{k_BT}},
\end{align}
\end{subequations}
where $\Delta \Phi^{(N)}_{i} \equiv \Phi_{i}^{(N)} - \Phi_{SLG}$, $\Phi_{i}^{(N)}$ and $\Phi_{SLG}$ is the Schottky barrier of $i$-stacking FLG and of single layer graphene Schottky interface, respectively. 
For simplicity, we assume that $\Delta \Phi^{(N)}_{i}$ is the same as the work function difference between monolayer graphene and FLG, which has a typical value of $\Delta \Phi^{(N)}_{i}\approx0.1$ eV \cite{aoki, filleter, yu, sque}. 
The temperature dependence of Eq. (\ref{ratio}) is shown in Fig. 3(d) for $N = (3, 4, 5, 7)$. 
At lower temperature regime $T \lesssim 400$ K, both $\bar{J}_{ABA}^{(N)}/\bar{J}_{Dirac}$ and $\bar{J}_{ABC}^{(N)}/\bar{J}_{Dirac}$ exhibit similar exponential-like growth with increasing $T$. 
Although Schottky devices are not typically operated at $T \gtrsim 400$ K, it is interesting to note that $\bar{J}_{ABA}^{(N)}/\bar{J}_{Dirac}$ and $\bar{J}_{ABC}^{(N)}/\bar{J}_{Dirac}$ exhibits contrasting high-temperature dependence. 
$\bar{J}_{ABA}^{(N)}/\bar{J}_{Dirac}$ maintains the exponential growth with increasing $T$ while $\bar{J}_{ABC}^{(N)}/\bar{J}_{Dirac}$ exhibits a gradual saturation. This can be traced back to the $\bar{J}_{ABC}^{(N)}/\bar{J}_{Dirac} \propto (1/T)^{2-2/N}$ dependence which balances out $e^{-\Delta \Phi^{(N)}_{ABA} /k_BT}$ at sufficiently high temperature.

The Richardson constant for  ABA-FLG and ABC-FLG can be defined, respectively, as $\mathcal{A}_{ABA}^{(N)} \equiv N\mathcal{B}$ and $\mathcal{A}_{ABC}^{(N)}\equiv \Gamma(2/N) (t_\perp/k_B)^{2-2/N} \mathcal{B}/N$. 
We plotted the Arrhenius plot with a representative FLG of $N=4$ in Fig. 3(e) using the actual scaling and the conventional $T^2$-scaling for comparisons. 
For $ABA$-FLG, $J_{ABA}^{(N=4)}/T^2$ are heavily \emph{up-shifted} by orders of magnitudes with respect to the actual scaling, $J_{ABA}^{(N=4)}/T^3$. 
This is contrary to $ABC$-FLG where $J_{ABC}^{(N=4)}/T^2$ is severely \emph{down-shifted} with respect to $J_{ABC}^{(N=4)}/T^{3/2}$. 
This immediately suggests that the Richardson constants extracted via the incorrect $T^2$-scaling can severely deviate from the actual values. 
For $ABA$-stacking, the Richardson constant fitted via $\log\left(J_{ABA}^{(N)}/T^2\right)$, i.e. $\mathcal{A}_{\text{fit}}$, yields a ratio of $\mathcal{A}_{\text{fit}}/\mathcal{A}^{(N)}_{ABA} = 870$ for all $N$, i.e. nearly $10^3$ overestimation. 
This extremely high overestimation remains approximately constant for all $N$ due to the universal $T^3$-scaling in $ABA$-FLG. 
In the inset of Fig. 3(e), $\mathcal{A}_{\text{fit}} / \mathcal{A}_{ABC}^{(N)}$ is plotted for $N$ up to 10 for $ABC$-FLG. 
The strong $N$-dependence is a consequence of the $T^{2/N+1}$-scaling. 
For $ABC$-trilayer graphene,  $\mathcal{A}_{\text{fit}}$ underestimates $\mathcal{A}_{ABC}^{(N=3)}$ by a factor of $\approx 0.1$. This underestimation becomes worsen and reaches $\mathcal{A}_{\text{fit}}/\mathcal{A}_{ABC}^{(N)} \approx 10^{-3}$ at $N=10$. 
In contrast, the extracted Schottky barrier height is not significantly influenced by different scaling relations due to the dominance of $e^{-\Phi/k_BT}$. Typically, $\Phi_{\text{fit}}/\Phi^{(N)}_{ABA} \approx 1.08$ and $\Phi_{\text{fit}}/\Phi^{(N)}_{ABC} \approx 0.93$ where $\Phi_{\text{fit}}$ is the Schottky barrier height fitted from the Arrhenius plot assuming a $T^2$ scaling. 
It should be noted that the good agreement between the $T^2$-fitted and the actual values of $\Phi$ \cite{yang, chen, yim, kim, mohammed, shivaraman, tongay, an, yu2} could misleadingly suggest the conventional $T^2$ model as a valid model for Schottky interfaces composed of non-parabolic energy dispersions. We summarize the main findings of this article in Table I.
\begin{widetext}
	\begin{center}
		\begin{table}
			\caption{Summary of the reverse saturation currents and the current-temperature scaling relations for Schottky transport model of non-parabolic energy dispersions. Note that $e^{-\frac{\Phi }{k_BT}}$ is omitted for simplicity. }
			\begin{tabular}{lll}
				\hline \hline
				Energy dispersion & Reverse saturation current & Scaling relation \\
				\hline
				$\displaystyle E_\parallel(1 + \gamma E_\parallel) = \frac{\hbar^2 k_\parallel^2}{2m} $ & $\bar{J}_{Kane} = \frac{g_{s,v} em k_B^2}{4\pi^2\hbar^3} \left(T^2 + 2\gamma k_BT^3\right)$ & $T^2 + 2\gamma k_BT^3$\\
				$E_\parallel(1+\gamma E_\parallel) = \frac{\hbar^2 k_x^2}{2m_x} + \frac{\hbar^2 k_y^2}{2m_y} $ & $\bar{J}_{anisotropy} = \frac{g_{s,v}e\sqrt{m_xm_y}}{4\pi^2 \hbar^3 k_B^2}\left( T^2 + 2k_B\gamma T^3 \right)$ & $T^2 + 2\gamma k_BT^3$ \\
				$E_\parallel =\frac{\hbar^2 k_\parallel^2}{2m} - \beta k_\parallel^4$ & $\bar{J}_{\alpha k^2 - \beta k^4} = \frac{g_{s,v}emk_B^2}{4\pi^2\hbar^3}\left(T^2 + \frac{8m^2\beta k_B}{\hbar^4}T^3\right)$ & $T^2 + \frac{8m^2\beta k_B }{\hbar^4}T^3$ \\
				$ E_{\parallel,n}  =  t_\perp \cos{\left(\frac{\pi n}{N+1}\right)} \pm \sqrt{\left(v_Fk\right)^2 + t_\perp^2 \cos^2{\left(\frac{\pi n}{N+1}\right)}} $ & $\bar{J}_{ABA}^{(N)} = N \times \frac{eg_{s,v} k_B^3}{ 4\pi^2 \hbar^3 v_F^2 } T^3 $ & $T^3$ \\
				$E_\parallel = \frac{ \left(\hbar v_F\right)^N }{ t_\perp^{N-1} }  k_\parallel^N$ & $\bar{J}_{ABC}^{(N)} = \frac{\Gamma\left(2/N\right)}{N} \left( \frac{t_\perp}{k_B} \right)^{2-\frac{2}{N}} \frac{eg_{s,v} k_B^3}{4\pi^2 \hbar^3 v_F^2 }   T^{ \frac{2}{N} + 1 }  $ & $T^{ \frac{2}{N} + 1 }$ \\
				\hline \hline
			\end{tabular}
		\end{table}
	\end{center}
\end{widetext}	

\section{Conclusion}

In summary, we found that the Schottky transport current in narrow-gap semiconductor and in FLG exhibits distinct forms of unconventional scaling relations. 
In practice, although the uncertainties of the Richardson constant extracted from an Arrhenius plot can be effectively reduced by using a Legendre polynomial fitting scheme as outlined in Ref. \cite{jensen}, using an incorrect scaling relation can still lead to a severe misinterpretation of the experimental data, yielding an extracted Richardson constant that differs with the actual value by two orders of magnitude. 
This highlights the importance of using the correct scaling relation in order to better understand the physics of Schottky devices based on novel materials with non-parabolic energy dispersions.

\section{Acknowledgment}
	This work is supported by Singapore Ministry of Education T2 grant (T2MOE1401) and USA AFOAR AOARD Grant (FA2386-14-1-4020). We thank Kelvin J. A. Ooi, M. Zubair and  S. J. Liang for useful discussions. 

\appendix

\section{Derivation of Kane-Schotkky model}

The non-parabolic Kane energy dispersion is given as \cite{kane, conwell, wu}
\begin{equation}
E_\parallel(1 + \gamma E_\parallel) = \frac{\hbar^2 k_\parallel^2}{2m},
\end{equation}
where $\gamma$ denotes the non-parabolicity of the band structure. The energy dispersion is a simple quadratic equation in $E_\parallel$ can be solved to obtain
\begin{equation}
E_\parallel = \frac{\sqrt{1+\frac{4\gamma\hbar^2 k_\parallel^2}{2m}} -1}{2\gamma}.
\end{equation}
This allows us to write down the following transformation:
\begin{equation}
kdk = \frac{m}{\hbar^2} \left(1 + 2\gamma E_k\right) dE_k.
\end{equation}
In this case, the function $n(E,E_\perp)$ is given as
\begin{eqnarray}
n(E,E_\perp) dE dE_\perp &=& g_{s,v} ev_\perp f(E)  \frac{d^3 k}{(2\pi)^3} \nonumber \\
&=& \frac{2\pi g_{s,v} e}{(2\pi)^3} f(E)  k_\parallel dk_\parallel \frac{1}{\hbar}  \frac{dE_\perp}{d k_\perp} d k_\perp \nonumber \\
&=& dE_\perp \frac{g_{s,v}em}{(2\pi)^2 \hbar^3}  \left[1+2\gamma (E - E_\perp) \right] f(E)dE, \nonumber \\
 && 
\end{eqnarray}
where $v_\perp = \hbar^{-1} dE_\perp / dk_\perp$ is the group velocity component along the emission direction. Assuming the Boltzmann statistic, i.e. $f(E) = e^{-\Phi/k_BT}$, the electron supply function can be written as
\begin{eqnarray}
N(E_\perp)dE_\perp &=& dE_\perp \frac{g_{s,v} em}{4\pi^2\hbar^3} \int_{E_\perp}^{\infty} \left[1 + 2\gamma (E-E_\perp)\right] f(E) dE \nonumber \\
&=& dE_\perp \frac{g_{s,v} em}{4\pi^2\hbar^3} (-k_BT) \left(1+2\gamma k_BT\right) e^{-\frac{E_\perp}{k_BT}}. \nonumber \\
 &&
\end{eqnarray}
This gives the Kane-Schottky diode equation as
\begin{equation}
\bar{J}_{Kane} = \frac{g_{s,v} em k_B^2}{4\pi^2\hbar^3} \left(T^2 + 2\gamma k_BT^3\right) e^{-\frac{\Phi }{k_BT}}.
\end{equation}
The Kane-Schotkky scaling relation exhibits a mixture of $T^2$ and $T^3$ behavior. In the extremely non-parabolic case, $\gamma \to \infty$, the energy dispersion becomes
\begin{equation}
E_\parallel = \lim_{\gamma\to\infty}\sqrt{\frac{2}{m\gamma}}\hbar k_\parallel.
\end{equation}
By defining $v_F = \sqrt{1/2m\gamma}$, the energy dispersion reduces the graphene's linear dispersion, i.e. $E_\parallel = \hbar v_Fk_\parallel$. Using the fact that $\gamma = (2mv_F^2)^{-1}$, we write
\begin{equation}
\bar{J}_{kane} = \frac{g_{s,v}ek_B^2}{4\pi^2\hbar^3 v_F^2} T^3 e^{-\frac{\Phi }{k_BT}},
\end{equation}
i.e. the modified Richardson law for graphene \cite{liang1}. In the other extreme case of $\gamma \to 0$, the conventional form of $\bar{J}\propto T^2$ can be obtained. In summary, 
\begin{equation}
\bar{J}_{Kane} =
\begin{cases}
\mathcal{A} T^2 e^{-\frac{\Phi }{k_BT}} & , \gamma\to 0;\\
\mathcal{B} T^3 e^{-\frac{\Phi }{k_BT}} & , \gamma \to \infty,
\end{cases}
\end{equation}
where $\mathcal{A} = \frac{g_{s,v}emk_B^2}{4\pi^2\hbar^3}$ and $\mathcal{B}=\frac{g_{s,v}ek_B^2}{4\pi^2\hbar^3 v_F^2}$.

\subsection{Anisotropic Kane's dispersion}

In the presence of anisotropy, Kane's non-parabolic energy dispersion can be written as
\begin{equation}
E_\parallel(1+\gamma E_\parallel) = \frac{\hbar^2 k_x^2}{2m_x} + \frac{\hbar^2 k_y^2}{2m_y} = \theta(\phi) \frac{\hbar^2 k^2}{2m_x} ,
\end{equation}
where $\theta(\phi) \equiv \cos^2\phi + (m_x/m_y) \sin^2\phi$, $(m_x, m_y)$ are the $x$-and $y$-direction electron effective mass and $\mathbf{k}_\parallel = (k_x, k_y) $. This gives
\begin{equation}
E_\parallel = \frac{\sqrt{1+\frac{4\gamma\theta(\phi)\hbar^2k_\parallel^2}{2m}}-1}{2\gamma},
\end{equation}
which leads to the $dk \to dE_\parallel$ relation of
\begin{equation}
k_\parallel dk_\parallel = \frac{1}{\theta(\phi)} \frac{m}{\hbar^2} \left(1+2\gamma E_\parallel\right) dE_\parallel.
\end{equation}
\begin{widetext}
The supply function density then becomes
\begin{equation}
n(E,E_\perp) dE dE_\perp = dE_\perp \frac{g_{s,v}em_x}{(2\pi)^3 \hbar^3}  \left[1+2\gamma (E - E_\perp) \right] f(E)dE \int_0^{2\pi} \left(\cos^2\phi + \frac{m_x}{m_y} \sin^2\phi\right)^{-1} d\phi .
\end{equation}
The angular integration has a closed form solution of 
\begin{equation}
\int_0^{2\pi} \left(\cos^2\phi + \frac{m_y}{m_x} \sin^2\phi\right)^{-1} d\phi = 2\pi\sqrt{\frac{m_y}{m_x}}.
\end{equation}
Therefore,
\begin{equation}
n(E,E_\perp) dE dE_\perp = dE_\perp \frac{g_{s,v}e\sqrt{m_xm_y}}{(2\pi)^2 \hbar^3}  \left[1+2\gamma (E - E_\perp) \right] f(E)dE.
\end{equation}
\end{widetext}
This is identical to the electron supply function of isotropy band except that the effective mass $m$ is replaced by the term $\sqrt{m_xm_y}$ in the pre-factor. Finally, the emission current density can be determined as
\begin{equation}
\bar{J}_{anisotropy} =\mathcal {C}\left( T^2 + 2k_B\gamma T^3 \right) e^{-\frac{\Phi}{k_BT}},
\end{equation}
where $\mathcal{C} \equiv  \frac{g_{s,v}e\sqrt{m_xm_y}}{4\pi^2 \hbar^3 k_B^2} $ is the modified \emph{anisotropic} Richardson constant.

\subsection{Parabolic dispersion with higher-order $k^4$ term}

Beyond the parabolic band approximation, a higher order term can be included to account for the non-parabolicity of the band structure at energy far away from the conduction band edge \cite{ruf}. In this case, the general form of the energy dispersion can be written as $E_\parallel =\alpha k_\parallel^2 - \beta k_\parallel^4$ where $\alpha = \hbar^2/2m$ and $\beta$ is a small correction factor. For this energy dispersion, we have
\begin{equation}
k_\parallel dk_\parallel = \frac{dE_\parallel}{2\sqrt{ \alpha^2 - 4\beta E_\parallel } }.
\end{equation}
This gives the supply function density of
\begin{equation}
n(E,E_\perp)dEdE_\perp = \frac{1}{8\pi^2\hbar}\frac{eg_{s,v}f(E)}{\sqrt{\alpha^2-4\beta(E-E_\perp)}} dEdE_\perp,
\end{equation}
and hence
\begin{equation}
N(E_\perp) dE_\perp = dE_\perp \frac{eg_{s,v}}{8\pi^2\hbar} \int_{E_\perp}^{\varepsilon_0+E_\perp} \frac{e^{-\frac{E}{k_BT}}}{\sqrt{\alpha^2 - 4\beta(E-E_\perp)}}.
\end{equation}
Note that the dispersion in Eq. (19) has a unphysical band turning at energy  $\varepsilon_0=\alpha^2/4\beta$. Hence, the upper limit of $\int dE_\parallel$ is set to $\varepsilon_0$. When converting $\int dE_\parallel \to \int dE$, the upper integration limit becomes $\varepsilon_0 + E_\perp$ since $E = E_\perp + E_\parallel$. The integral can be solved analytically as
\begin{equation}
N(E_\perp) dE_\perp = \frac{dE_\perp}{2} \sqrt{\frac{\pi}{\beta}} \left(k_BT\right)^{1/2} e^{-\frac{\varepsilon_0}{k_BT}} erfi\left(\sqrt{\frac{\varepsilon_0}{k_BT}}\right) e^{-\frac{E_\perp }{k_BT}},
\end{equation}
where $erfi(x)$ is the \emph{imaginary error function}. Finally, we obtain
\begin{equation}
\bar{J}_{\alpha k^2 - \beta k^4} = \frac{e}{2}\sqrt{\frac{\pi}{\beta}} \left( k_BT \right)^{3/2} e^{-\frac{\varepsilon_0}{k_BT}} erfi\left(\sqrt{\frac{\varepsilon_0}{k_BT}}\right) e^{-\frac{\Phi}{k_BT}}.
\end{equation}
Using the identity $e^{-x^2} erfi(x) = \frac{2}{\sqrt{\pi}} \mathcal{D}_+(x)$ where $\mathcal{D}_+(x) \equiv e^{-x^2} \int^x_0 e^{t^2} dt$ is the Dawson integral, the current density can be re-written as
\begin{equation}
\bar{J}_{\alpha k^2 - \beta k^4} = \frac{g_{s,v}e}{8\pi^2\hbar \sqrt{\beta}} \left(k_BT\right)^{3/2} \mathcal{D}_+\left(\sqrt{\frac{\varepsilon_0}{k_BT}}\right)e^{-\frac{\Phi }{k_BT}}
\end{equation}.
\begin{widetext}
In the limit of $\varepsilon_0>>k_BT$, 
\begin{equation}
\bar{J}_{\alpha k^2 - \beta k^4} = \frac{g_{s,v}e}{8\pi^2\hbar \sqrt{\beta}} \left(k_BT\right)^{3/2} \left[ \frac{1}{2}\left(\frac{k_BT}{\varepsilon_0}\right)^{1/2}  + \frac{1}{4} \left(\frac{k_BT}{\varepsilon_0}\right)^{3/2}\right]e^{-\frac{\Phi }{k_BT}} ,
\end{equation}
where the identity of $\mathcal{D}_+(x) \approx 1/2x + 1/4x^3 + \cdots$ for large $x$ has been used. Replacing $\alpha = \hbar^2/2m$, we obtain the final form of
\begin{equation}
\bar{J}_{\alpha k^2 - \beta k^4} = \frac{g_{s,v}em^*}{4\pi^2\hbar^3}\left[\left(k_BT\right)^2 + \frac{8m^2\beta}{\hbar^4}\left(k_BT\right)^3\right]e^{-\frac{\Phi }{k_BT}} .
\end{equation}
\end{widetext}

\section{Schottky model in few-layer graphene}

We now derive the reverse saturation current in FLG. FLG can be stacked according to two stacking orders: (i) Bernal $ABA$-stacking; and (ii) rhombohedral $ABC$-stacking \cite{bao, zhang_F, min, koshino}. Experimentally, it was shown that the ABC-staking made up of 15\% of the total area of mechanically exofoliated tri-and tetra-layer graphene \cite{li}. For chemically grown graphene multilayer in SiC substarte, ABC-stacking is the dominant configuration \cite{norimatsu}. For completeness, FLG of both $ABA$-and $ABC$-stacking are considered. 

\subsection{$ABA$-stacked few-layer graphene}

For ABA-stacked FLG, we rewrite the energy dispersion in Eq. (11) of the main text as
\begin{equation}
E_{\parallel,n} = \frac{1 \pm \sqrt{ \gamma_{N,n}^2 \hbar^2 v_F^2k_{\parallel,n}^2  + 1} }{\gamma_{N,n}},
\end{equation}
where $\gamma_{N,n} \equiv \left[t_\perp \cos{\left(\frac{\pi n}{N+1} \right)}\right]^{-1}$. Hence,
\begin{equation}
k_{\parallel,n} dk_{\parallel,n} = \frac{\gamma_{N,n} E_{\parallel,n} - 1}{\gamma_{N,n} \hbar^2v_F^2} dE_{\parallel,n}.
\end{equation}
The supply function due to electrons from $n$-subband is given as
\begin{equation}
N^{(n)}(E_\perp) dE_\perp = dE_\perp \frac{eg_{s,v}}{4\pi^2\hbar^3 v_F^2} \left[ (k_BT)^3 - \frac{(k_BT)^2}{\gamma_{N,n} } \right] e^{-\frac{E_\perp}{k_BT}}.
\end{equation}
The reverse saturation current density can be calculated as
\begin{widetext}
\begin{eqnarray}
\bar{J}^{(N)}_{ABA} &=& \sum_{n=1}^{N} \int_{\Phi^{(N)}_{ABA}}^{\infty} N^{(n)}(E_\perp) dE_\perp \nonumber \\
&=& \frac{eg_{s,v}}{4\pi^2\hbar^3v_F^2} \sum_{n=1}^{N} \left[ (k_BT)^3 - t_\perp \cos{\left(\frac{\pi n}{N+1}\right)} (k_BT)^2 \right] e^{-\frac{\Phi^{(N)}_{ABA}}{k_BT}}.
\end{eqnarray}
\end{widetext}
Note that the cosine term in the square bracket follows the following identity
\begin{equation}
\cos{\left( \frac{j\pi}{k} \right)} = -\cos{\left( \frac{N-j}{k} \pi \right)},
\end{equation}
where $j$ is a positive integer with $2j \neq N$ and $j \textless N$. Therefore, the summation over all $n$ results in the mutual-cancellation of the $T^2$ terms in the square bracket in the second line of Eq. (B4). This gives the \emph{total} reverse saturation current of
\begin{equation}
\bar{J}_{ABA}^{(N)} = N \times \frac{eg_{s,v} k_B^3}{ 4\pi^2 \hbar^3 v_F^2 } T^3 e^{-\frac{\Phi^{(N)}_{ABA} }{k_BT}} .
\end{equation}

\subsection{$ABC$-stacked few-layer graphene}

For ABC-stacked $N$-layer graphene with $N\geq 2$, the low energy two-band effective Hamiltonian can written as \cite{min}
\begin{equation}
\hat{\mathcal{H}}_{\mathbf{k}} =  - \frac{ \left(\hbar v_F\right)^N }{ t_\perp^{N-1} } 
\begin{pmatrix}
0 & k_-^N \\
k_+^N & 0
\end{pmatrix},
\end{equation}
where $k_\pm = k_x \pm ik_y$. The basis of $\hat{\mathcal{H}}_{\mathbf{k}}$ is composed of the sublattices in the outermost layers, i.e. $(\phi_{A_1}, \phi_{B_N})^T$, since they are responsible for the low-energy dynamics. By diagonalizing $\hat{\mathcal{H}}_{\mathbf{k}}$, the energy dispersion is found to be $E_\parallel = \alpha_N k_\parallel^N$ where $\alpha_N \equiv \left( \hbar v_F \right)^{N-1} / t_\perp^{ N-1 }$. Similarly, the following relation can be determined
\begin{equation}
k_\parallel dk_\parallel = \frac{1}{N \alpha_N} \left(\frac{E_\parallel}{\alpha_N} \right)^{\frac{2}{N} - 1}.
\end{equation}
Similarly, the electron supply function can be written as
\begin{widetext}
\begin{equation}
N(E_\perp) dE_\perp = d E_\perp \frac{g_{s,v} e}{4\pi^2 \hbar N \alpha_N} \int^\infty_{E_\perp} dE \left(\frac{E-E_\perp}{\alpha_N}\right)^{\frac{2}{N} - 1} e^{-\frac{E}{k_BT}}.
\end{equation}
The integral can be analytically solved in terms of an incomplete gamma function, i.e.
\begin{equation}
\int^\infty_{E_\perp} dE \left(\frac{E-E_\perp}{\alpha_N}\right)^{\frac{2}{N} - 1} e^{-\frac{E}{k_BT}} = -\alpha_N^{1 - \frac{2}{N}} \left( k_BT \right)^{\frac{2}{N}} \Gamma\left(\frac{2}{N}\right) e^{-\frac{E_\perp }{k_BT}}
\end{equation}.
Finally, the current density is found to be
\begin{equation}
\bar{J}_{ABC}^{(N)} = \frac{eg_{s,v} k_B^3}{4\pi^2 \hbar^3 v_F^2 } \frac{ \left( t_\perp k_B \right)^{2-\frac{2}{N}} }{N}\Gamma\left(\frac{2}{N}\right) T^{ \frac{2}{N} + 1 }  e^{-\frac{\Phi_{ABC}^{(N)} }{k_BT}}.
\end{equation}

\end{widetext}

\end{document}